\documentstyle[12pt]{article}
\title{Gap nodes in the superconducting phase of the itinerant
 ferromagnet $UGe_2$.}
\author{I.A.Fomin,\\
P. L. Kapitza Institute for Physical Problems,\\
Kosygina 2, 117334 Moscow, Russia}
\date{ }
\begin{document}

\maketitle
\begin{abstract}
For the UGe$_2$ ferromagnetic superconductor the forms of the order parameter
admitted by the crystal symmetry within the strong spin-orbit coupling scheme
are written down. For each of the two possible phases existence of
 gap nodes required by symmetry is discussed and the nodes are found.
Some consequences of presence of the nodes, which may be useful for
 experimental identification of the phases are discussed as well.
 \end{abstract}
 
 \section {Introduction}
  The itinerant ferromagnet UGe$_2$ becomes superconducting in a pressure
  interval $11<P_c<16 kbar$ at a temperature below $0.8K$ \cite{sax,huxley}.
  This temperature is much smaller then the Curie temperature $T_c$ for the
 same pressure. The estimated splitting of (quasi)spin-up and spin-down Fermi
 surfaces is 2 - 3 orders of magnitude greater then the superconducting gap.
 This condition rules out the possibility of the spin-singlet Cooper pairing.
 At a triplet pairing the order parameter is a complex vector function
 ${\bf d}({\bf k})$.  As a consequence of time reversal symmetry breaking
 superconducting phases of a ferromagnet generally speaking are nonunitary,
 i.e.  ${\bf  d}$ and ${\bf d}^{\star}$ are not proportional to each other,
 or  ${\bf d}\times{\bf d}^{\star}\ne 0$.
 Possible forms of the order parameters for the superconducting phases emerging
 continuously from a given normal phase can be classified if the group of
 symmetry of the normal phase is known \cite{volovik}. An attempt of such
 classification for UGe$_2$ has been made in a ref.\cite{fomin} . It has been
 observed that  the magnetic group of an orthorombic ferromagnetic
 crystal  $D_2(C_2)$ is isomorphic to $D_2$ and the basis functions of
 four different representations of the group $D_2$:  А, B$_1$, B$_2$, B$_3$
 were suggested as possible forms of the order parameter.
  It was shown later  (ref. \cite{mineev}) that not all of the suggested
  functions are independent.  Because of the specific rules for the
 multiplication of antiunitary elements of magnetic groups the basis functions
 corresponding to the representations A and B$_1$ are equivalent i.e. are
 transformed according to one corepresentation of the magnetic group.  Two
 other functions are also mutually equivalent. As a result there are only two
  essentially different types of symmetry of the superconducting order
  parameter.

 For an experimental identification of a type of the order parameter,
 which is realised in UGe$_2$  the existence and of gap nodes
 and their character for each of the phases are of importance.
 The aim of the present paper is to find the nodes. The following arguments
 apply  as well to another orthorombic itinerant superconducting ferromagnet
 -- URhGe  \cite{aoki}.

\section {Basis functions}
Let us find first a general form of the functions ${\bf \Psi}_A$ and
${\bf \Psi}_B$, which are transformed under two different corepresentations
A and B of magnetic group $D_2(C_2)$. The group $D_2(C_2)$ contains four
operators. Two of them -- the unity operator $E$ and the rotation for an angle
$\pi$ around z-direction $C_2^z$ are unitary. Two others -- $RC_2^x$ and
$RC_2^y$ contain time reversal $R$ and are nonunitary.
Corepresentations are formed by matrices $G_1$ and $G_z$ corresponding to
the unitary operators and  $F_x$, $F_y$ -- to the nonunitary. In the present
case corepresentations are one-dimensional and the matrices $G_1, G_z, F_x, F_y$
are just complex numbers. According to the rules of multiplication of matrices
forming  corepresentations \cite{wigner} they satisfy the following equations
$G_z^2=G_1, F_x\cdot F_x^{\star}=G_1, F_y\cdot F_y^{\star}=G_1, F_x\cdot
F_y^{\star}=G_z $.
These equations have two solutions, giving rise to two corepresentations.
One of them (referred as A) has a form:
$$
G_1=1; G_z=1;  F_x=e^{2i\phi}; F_y=e^{2i\phi}.  \eqno(1)
$$

The other -- B a form:
$$
G_1=1; G_z=-1; F_x=e^{2i\phi}; F_y=-e^{2i\phi}, \eqno(2)
$$

where $\phi$ is a real scalar.
Let us write down ${\bf \Psi}_A$ in a form adopted for the strong spin-orbit
  coupling scheme \cite{volovik}:

  $$
  {\bf \Psi}_A={\bf x}f_x({\bf k})+{\bf y}f_y({\bf k})+
  {\bf z}f_z({\bf k}),                              \eqno(3)
  $$

  where {\bf x}, {\bf y}, {\bf z}- unit vectors directed along the symmetry axes {\bf b}, {\bf c}, {\bf a}. In
 the ferromagnetic phase {\bf a} is an easy magnetization axis. All functions
$f_{x,y,z}({\bf k})$ are odd with respect to {\bf k}, i.e.  $f_x(-{\bf
 k})=-f_x({\bf k})$ etc..  ${\bf \Psi}_A$ under the action of operators $E;
 C_2^z; RC_2^x; RC_2^y$ is multiplied by the numbers specified by the eq. (1).
That imposes constraints on the functions $f_x({\bf k}), f_y({\bf k}), f_z({\bf
k})$.  Since all operators in question are linear or antilinear the constraints
are imposed separately on each function   $f_x, f_y, f_z$.  For $f_x({\bf k})$
it is
$$
\begin{array}{rcl} f_x(-k_x,-k_y,k_z)=-f_x(k_x,k_y,k_z); \\

f_x^{\star}(k_x,-k_y,-k_z)  = e^{2i\phi} f_x(k_x,k_y,k_z);\\

f_x^{\star}(-k_x,k_y,-k_z)  = e^{2i\phi} f_x(k_x,k_y,k_z). \\

\end{array}
$$
 The constraints for $f_y({\bf k})$ are obtained by interchange of indices $x$
 and $y$.
 The constraints for $f_z({\bf k})$ are:

$$
\begin{array}{rcl}

f_z(-k_x,-k_y,k_z)=f_z(k_x,k_y,k_z);\\

f_z^{\star}(k_x,-k_y,-k_z)  = - e^{2i\phi} f_z(k_x,k_y,k_z);\\

f_z^{\star}(-k_x,k_y,-k_z)  = - e^{2i\phi} f_z(k_x,k_y,k_z).\\

\end{array}
$$

 The above conditions are satisfied by the following function

$$
  {\bf \Psi}_A=
  e^{-i\phi_A} \{{\bf\hat x}k_x(a_{11}+ik_xk_ya_{10})+
  {\bf\hat y}k_y(a_{22}+ik_xk_ya_{20})+
  {\bf\hat z}k_z(a_{33}+ik_xk_ya_{30})\},       \eqno(4)
$$

 where $\phi_A$, $a_{11}$,... are real functions of $k_x^2,k_y^2,k_z^2$.
 ${\bf \Psi}_A$ defined by eq. (4) differs from that given by eq.(2)
 of ref.\cite{fomin} by the phase factor $e^{-i\phi_A} $.
 When $\phi_A=\pi/2 $, ${\bf \Psi}_A$ defined by eq. (4)
 can be cast in a form coinciding with  ${\bf \Psi}_{B_1}$ of ref.\cite{fomin}.

 For the corepresentation B the constraints for $f_x({\bf k})$ are:
$$
\begin{array}{rcl}
f_x(-k_x,-k_y,k_z)=-f_x(k_x,k_y,k_z), \\

f_x^{\star}(k_x,-k_y,-k_z)  = e^{2i\phi} f_x(k_x,k_y,k_z),\\

f_x^{\star}(-k_x,k_y,-k_z)  = e^{2i\phi} f_x(k_x,k_y,k_z), \\

\end{array}
$$

and for $f_z({\bf k})$:

$$
\begin{array}{rcl}

f_z(-k_x,-k_y,k_z)=f_z(k_x,k_y,k_z);\\

f_z^{\star}(k_x,-k_y,-k_z)  = - e^{2i\phi} f_z(k_x,k_y,k_z);\\

f_z^{\star}(-k_x,k_y,-k_z)  = - e^{2i\phi} f_z(k_x,k_y,k_z).\\

\end{array}
$$

A general form of the basis function for the corepresentation B reads as:

 $$
  {\bf \Psi}_B=
  e^{-i\phi_B}\{{\bf\hat x}k_z(b_{13}+ik_xk_yb_{10})+
  {\bf\hat y}k_z(ib_{23}+k_xk_yb_{20})+
  {\bf\hat z}k_x(b_{31}+ik_xk_yb_{30})\}.       \eqno(5)
 $$
With a suitable choice of the phase factor $e^{-i\phi_B}$ it can be
transformed either in  ${\bf \Psi}_{B_3}$ or in ${\bf \Psi}_{B_4}$  of ref.
\cite{fomin}.

\section {Nodes}
At a triplet Cooper pairing gaps in the spectra of one-particle excitations
are determined \cite{woelf}  by the eigenvalues of the matrix
$$
 (\Delta_k\Delta^{\dagger}_k)_{\alpha\beta}=
 {\bf d(k)}\cdot{\bf d^*(k)}\delta_{\alpha\beta} +
 i[{\bf d(k)}\times{\bf d^*(k)}]{\bf \sigma}_{\alpha\beta}. \eqno(6)
$$
For nonunitary phases this matrix has two different eigenvalues. Each of them
gives a square of a gap for one projection of spin.
In terms of the real and imaginary parts of ${\bf d(k)}$, i.e.
${\bf d(k)}={\bf d_1(k)}+i{\bf d_2(k)}$ the eigenvalues of
$(\Delta_k\Delta^{\dagger}_k)_{\alpha\beta}$ read as
$|\Delta_{1,2}|^2={\bf d_1(k)}^2+{\bf d_2(k)}^2\pm 2|{\bf d(k)}
\times{\bf d^*(k)}|$.
The gap turns to zero for both projections of spin when
${\bf d_1(k)}=0$ and ${\bf d_2(k)}=0$; and only for one of the two projections
when $|{\bf d_1(k)}|=|{\bf d_2(k)}|$ and $|{\bf d_1(k)}|\perp|{\bf d_2(k)}|$.
Direct check shows that for a general form of unknown functions
 $a_{11},a_{10}...,b_{13},b_{10}...$ in the expressions (4) and (5) for both
 types of the order parameter ${\bf \Psi}_A$ and ${\bf \Psi}_B$ there are no
 nodes in the gaps. It means that the nodes are not required by symmetry.
 We have not used yet the fact that the splitting of
 spin-up and spin-down Fermi surfaces in UGe$_2$ and in URhGe is large.

The splitting of Fermi-surfaces suppresses the pairing amplitude  for
quasiparticles with different spin projections.
For a singlet Cooper pairing superconductivity is completely destroyed
at a splitting $2I>\sqrt{2}\Delta_0$, where $\Delta_0$ is a gap at zero
temperature and zero splitting \cite{abrikos}.

The formal reason for suppression of the pairing is the smearing of the
singularity of the scattering amplitude of two quasiparticles with the opposite
momenta and opposite spin projections at the polarization. When the
polarization is absent the scattering amplitude in a second order on
interaction has a singular contribution of a form: $\ln{(\omega_D/\Delta_0)}$.
At the polarization the splitting of two Fermi surfaces  $2I$  occurs and
the singular part turns into  $\ln{(\omega_D/I)}$. When $I\gg\Delta_0$
this contribution can be included in a regular part of the scattering
amplitude.  Transition from $\Delta_{\uparrow\downarrow}\ne 0$  to
 $\Delta_{\uparrow\downarrow}=0$ must take place at
$I\sim\Delta_0\sim T_s$.  Both in  UGe$_2$ and in  URhGe the condition $I\gg T_s$
is well satisfied, so it is safe to assume that
$\Delta_{\uparrow\downarrow}=0$ for these compounds. This is equivalent to
the condition $d_z({\bf k})=0$.
With that constraint two types of the order parameter acquire the form:
$$
{\bf \Psi}_A= e^{-i\phi_A} \{{\bf\hat
x}k_x(a_{11}+ik_xk_ya_{10})+ {\bf\hat y}k_y(a_{22}+ik_xk_ya_{20})\}, \eqno(7)
$$

$$
{\bf \Psi}_B= e^{-i\phi_B}\{{\bf\hat x}k_z(b_{13}+ik_xk_yb_{10})+
{\bf\hat y}k_z(ib_{23}+k_xk_yb_{20})\}.  \eqno(8)
$$

 ${\bf \Psi}_A$ together with the gaps on both Fermi surfaces turns to zero
 at the points  $k_x=0, k_y=0$. These are symmetry nodes. To see it consider
  ${\bf \Psi}_A(0,0,k_z)$ and apply to this function operation  $С_2^z$.
 At the strength of eq. (1)
 $С_2^z{\bf \Psi}_A(0,0,k_z)={\bf \Psi}_A(0,0,k_z)$.
 On the other hand from the definition of  $С_2^z$ :
 $$
 С_2^z{\bf \Psi}_A(0,0,k_z)=-{\bf x}f_x(0,0,k_z)-{\bf y}f_y(0,0,k_z)=
 -{\bf \Psi}_A(0,0,k_z)                          \eqno(9)
 $$
 Comparing two results we arrive at ${\bf \Psi}_A(0,0,k_z)=0$.

 In a similar way eq. (8) indicates that $ {\bf \Psi}_B$ turns to zero
 on a line $k_z=0$. These nodes are also required by symmetry since
 $$
 С_2^z{\bf \Psi}_B(k_x,k_y,0)=
 -{\bf x}f_x(-k_x,-k_y,0)-{\bf y}f_y(-k_x,-k_y,0)=
 {\bf \Psi}_B(k_x,k_y,0)                          \eqno(9)
 $$

 On the other hand from eq. (2)
 $С_2^z{\bf \Psi}_B(k_x,k_y,0)=-{\bf \Psi}_B(k_x,k_y,0)$, i.e.
 ${\bf \Psi}_B(k_x,k_y,0)=0$.

\section {Discussion}
The above argument shows that two possible superconducting phases of
UGe$_2$ differ in a character and position of the gap nodes.
For the A-type phase (eq.(7))  these are isolated nodes at points of intersection of
the Fermi surfaces with the direction of easy magnetization axis. For the
B-type phase (eq.(8)) these are lines of nodes on equators of the Fermi
surfaces which are perpendicular to the above mentioned axis.
The nodes are giving rise to power-law dependencies of thermodynamic quantities
on temperature at  $T\ll T_s$.  The exponents depend on a character of the
nodes. Investigation of the power-law dependencies of thermodynamic quantities
 is a standard tool for identification of unconventional superconducting phases
 \cite{samo}. Let us point out some specifics of the expected low temperature
properties of UGe$_2$, stemming from its magnetic polarization.
1)The values of gaps are generally speaking  different for different spin
projections. At a large splitting of two Fermi surfaces one can expect that
this difference can be large as well, for example
$\Delta_{\downarrow}\ll \Delta_{\uparrow}$. Then in a temperature interval
$\Delta_{\downarrow}\ll T< T_s$ the smaller gap practically does not influence
temperature dependencies of the thermodynamic quantities and the contribution of
spin-down quasiparticles to thermodynamics will be practically that of the
normal phase.
2)Magnetic field, induced by spontaneous magnetization
 ${\bf H}_M=4\pi{\bf M}$ in UGe$_2$ is  $H_M\sim$1 кое.
This is much greater then estimated $H_{C1}$ for that compound.
It means that UGe$_2$ is in a mixed state (or in the spontaneous flux phase
\cite {varma}). Combination of the vortices with the line of nodes oriented
perpendicular to the axes of vortices according to ref. \cite{gvol} gives rise
to a finite density of states on a Fermi level, which in its turn renders a
linear temperature dependence of specific heat at low temperatures with a
coefficient proportional to a square root of the field:
$c_s\sim c_n\sqrt{H_M/H_{c2}}$.  For fields much smaller then
 $H_{c2}$ this contribution to the specific heat is more important then the
 contribution of the bound electron states in the cores of the vortices.
 According to the present analysis a contribution of the discussed type is
expected in  the B-type phase, but not in the A-type.
One can conclude that the expected difference in the low temperature properties
of A and B-type phases can be used for identification of
superconducting phases realized in  UGe$_2$ and in URhGe.

  Part of this work was completed in CEA Grenoble. I am grateful to
  J.Flouquet for hospitality in this research center and for stimulating
  discussions, to Universite Joseph Fourier for the financial support of my stay
  in Grenoble, to V.P.Mineev and A.Huxley for valuable comments and
  discussions. This work was also supported by RFFS ander grant  010216714.

  \end{document}